\documentclass[12pt]{article}

\usepackage[cp1251]{inputenc}

\begin{document}
\normalsize

\title{\bf Superradiance on the Landau levels and the problem of  power of
 decameter radiation of Jupiter}

\author{{P.I.Fomin$^{*\,1}$, A.P.Fomina$^1$ and V.N.Malnev$^2$}
\bigskip \\
{\it $^1$Bogolyubov Institute for Theoretical Physics, Kiev 03143,
Ukraine} \\ {\it $^2$ Shevchenko Kiev National University, Kiev
03022, Ukraine}}

\date{\empty}
\maketitle

\begin{center}
%\bigskip
(February 20, 2002)
\bigskip
\end{center}

\begin{abstract}
We determine the conditions of formation of spontaneous
polarization phase transition to the superradiance regime in the
inverted system of nonrelativistic electrons on equidistant Landau
levels in rarefied magnetized plasma. The possibility of
realization of such conditions in the lower Jupiter magnetosphere
is shown. The effect of cyclotron superradiance on the Landau
levels gives a key to  interpretation of the nature of superpower
radioemission  of the Jupiter-Io system.
\end{abstract}

\bigskip
PACS number(s): 42.50.Fx, 41.60.-m

\vfill
$^*$ E-mail: pfomin@bitp.kiev.ua

\newpage

%\Large

\sloppy

\begin{center}
\section{Introduction.}
\end{center}

The phenomenon of superradiance (SR) was considered for the first time in the
known paper by Dicke [1] in the example of a two-level model. At present, a
significant number of publications are devoted to its study (see for example,
reviews [2,3,4]), but, as noted by many authors, the theme is far from being
exhausted, and many interesting and physically important questions and
situations still remain to be investigated. One of them is the question of the
possibility of realization of SR in systems with equidistant levels in the
presence of inversion. The system of fast electrons in a homogeneous magnetic
field, for which the transversal movement (rotation) energy spectrum is
described by the known Landau levels
\begin{equation}
E_{\perp}=n \hbar \omega, \quad n=0,1,2,\dots \, ,
\end{equation}
\begin{equation}
\omega=\frac{eH}{mc} \, ,
  \end{equation}
is the most easily realizable and important of such systems [5]. The question
on the possibility and conditions of SR formation in such systems, besides
general physical interest, is also of big astrophysical interest because the
inversion on the Landau levels is easily achieved in rarefied magnetized plasma
of active space objects in the presence of bunches of accelerated electrons if
their initial velocities are directed at some angle to the magnetic field, and
so, besides the longitudinal energy $E _ {\parallel}$, they also possess
significant transversal energy $E _ {\perp}$. For example, in the Jupiter
magnetosphere [6,7], the observed superpower sporadic nonthermal decameter
radiation of the Jupiter-Io system with brightness temperature up to
$10^{15}$--$10^{17}$~K can serve as an evidence of the possible realization of
such situation and generation of SR.

In the present work, we investigate the question about the possibility and
conditions of SR formation in the inverted system of electrons on high Landau
levels (1) with $n \gg 1$ and the possibility of realization  of these
conditions in the lower Jupiter magnetosphere. As it is known, for the
generation of the induced coherent radiation in systems like masers, the
equidistancy of energy levels is an obstacle because of the specific
competition of radiation and absorption processes in this case. In the case of
SR, we deal not with induced but with spontaneous radiation, and here, as we
shall see, equidistancy of energy levels appears to be of advantage. This is,
first of all, because the SR regime is usually realized in open finite systems
without mirrors, where radiation leaves the active volume of generation quickly
enough, having practically no time to get in the absorption regime [3]. Second,
in this case, all the inverted electrons occupy, as a rule, not one level, but
some significant interval of high levels $ \triangle n$  $(n \gg \triangle n
\gg 1) $, and because of equidistancy, all of them radiate the same mode on
frequency (2), and also, as we shall see, at the same rate independent of the
initial energy.

The phenomenon of SR arises when in ``coherence domains", with sizes $R_0$
smaller then the wavelength $ \lambda $, all $N_0 $ radiating dipoles gradually
during radiation become aligned in one direction due to the dipole-dipole
interaction between them in the ``near zone" $ (R_0 \ll \lambda) $, so that, as
a result, the total dipole of the domain $ \overrightarrow {D} $ turns out to
be $N_0 $ times larger then  the elementary dipole $ \overrightarrow {d} $.
Therefore, the intensity of the collective dipole radiation becomes
proportional to $N^2_0 $, and not $N_0 $ as in the case of radiation of
uncorrelated dipoles. This is described in Sec.~2.

The transition to such a correlated polarized state is similar to
the phase transition in magnetics or ferroelectrics, and for its
description it is convenient to use the Weiss method of mean
self-consistent field [8]. This theory is developed in Sec.~3. We
note that the phase transition under consideration is a
nonequilibrium one, and it has all the features of the
self-organization phenomena in dissipative systems. In Sec.~4, we
discuss the key role of the SR effect for the interpretation of
the observable power of the decameter radiation of the Jupiter-Io
system.
\bigskip

\section{ Cyclotron superradiance on Landau levels.}

For large $n$, levels (1) correspond to quantum states with wave functions
localized near classical Larmor orbits with radii $r _ {L} =V _ {\perp}/\omega
$ [5]:
\begin{equation}%3
r_n =\sqrt {\frac {c\hbar} {eH} (2n +1)} =
  \sqrt {\frac {2 n \hbar\omega} {m\omega^2} \left(1 +\frac {1} {2n}\right)}
  \approx \sqrt {\frac {2E _ {\perp}} {m \omega^2}} =
  \frac {V _ {\perp}} {\omega} =r_L \, .
\end{equation}
In this connection, we can proceed to (quasi)classical description of such
states and transitions between them. In a classical limit, in a coordinate
system in which the longitudinal movement is absent, these orbits  are given by
\begin{equation}%4
\overrightarrow{r}_{\perp}(t)=r_{L} \{ \cos(\omega t+\alpha), \ \sin (\omega
t+\alpha), \ 0 \} \, ,
\end{equation}
where the cartesian components of the vector $ \overrightarrow {r} _ {\bot} $
are written in braces.

Being initially inverted on high levels, electrons start to fall down step by
step on the ladder of states (1), radiating quanta with frequency (2). The
differential (angular) and integral intensities of the dipole radiation of one
electron in the classical limit are described by the known formulas [9]
\begin{equation}%5
\frac{dI}{d\Omega}=\frac{[\ddot{\overrightarrow{d}}\times
  \overrightarrow{n}_{k}]}{4 \pi c^{3}} , \quad
  \overrightarrow{n}_{k}\equiv \frac{\overrightarrow{k}}{k} ;
\end{equation}
\begin{equation}%6
  I=\frac{2e^{2}\omega^{2} V^{2}_{\perp}}{3c^{3}}
  =\frac{4e^{2}\omega^{2}}{3mc^{3 }}\;
  E_{\perp}.
\end{equation}
Equation (6) implies the following evolution law of the electron energy:
\begin{equation}%7
  \frac{dE_{\perp}(t)}{dt}=-I=-\frac{E_{\perp}}{\tau},
  \quad \tau =\frac{3mc^3}{4e^2\omega^{2}} ;
\end{equation}
\begin{equation}%8
E_{\perp}(t)=E_{\perp}(0) \; \exp(-t/\tau).
\end{equation}
We see that the radiation time $ \tau $ does not depend on $ E _ {\perp} $,
i.e., electrons with different initial energies [within the limits of
dispersion $ \triangle E _ {\perp} (0) \ll E _ {\perp} (0) $] will fall down at
the same rate. It is easy to see that the dispersion of energy will also
decrease at the same rate,
\begin{equation}%9
 \triangle
 E_{\perp}(t)=\triangle E_{\perp}(0) \; \exp(-t/\tau),
\end{equation}
so that the following relation is satisfied:
\begin{equation}%10
\triangle
 E_{\perp}(t)/E_{\perp}(t) = {\rm const} \ll 1.
\end{equation}
This allows one to judge about the evolution of the whole collective of
electrons by the evolution of their average energy and other quantities.
Therefore, for simplicity, in what follows, by the symbols $ E_{\perp}$,
$V_{\perp}$, and $ d_{0}=er_{L}=eV_{\perp} / \omega $ we denote the
corresponding values averaged over the ensemble of inverted electrons.

One can partition the total volume $V $,  occupied by $ N $ inverted electrons,
into subvolumes $V _ {coh} $, or ``coherence areas" describing the elementary
dipoles the sizes $ R _ {0} $ of which are smaller than the radiated wavelength
$ \lambda $ but are bigger than the radii of the Larmor orbits (3):
\begin{equation}%11
  r_L \ll R_{0} \ll \lambda.
\end{equation}
Thus,  a large enough number of  $N _ {0} $ radiating dipoles will be in one
volume $V _ {coh} $:
\begin{equation}%12
  n_{e}V=N \gg N_{0}=n_{e} V_{coh} \gg 1.
\end{equation}
Let us consider the total dipole moment of such a subsystem
\begin{equation}%13
  \overrightarrow{D}(t)=\sum\limits_{j=1}^{N_{0}} \overrightarrow{d_{j}}(t)
  =e\sum\limits_{j=1}^{N_{0}}
  \overrightarrow{r_{j }}_{\perp}(t).
\end{equation}

Similarly to (5) for one electron, the collective  dipole radiation of our
subsystem will be described by the formula
\begin{equation}%14
  dI=\frac{[\ddot{\overrightarrow{D}}\times
  \overrightarrow{n_{k}}]^2}{c^3}\frac{d\Omega}{4 \pi}=\frac{\omega^4}{c^3}
  \left[\overrightarrow{D}^2-(\overrightarrow{D}\cdot
  \overrightarrow{n_{k}})^2 \right] \frac{d\Omega}{4\pi}.
\end{equation}
By substituting (13) and (4) into this formula and averaging over the period
$T=2\pi/\omega $, we have
$$
  \left\langle \cos(\omega t + \alpha_i) \cos(\omega t +
  \alpha_j) \right\rangle  = \cos(\alpha_i- \alpha_j)/2,
$$
$$
\left\langle \sin(\omega t + \alpha_i) \sin(\omega t +
  \alpha_j) \right\rangle  = \cos(\alpha_i- \alpha_j)/2,
$$
\begin{equation}%15
  \left\langle \cos(\omega t + \alpha_i) \sin(\omega t +
  \alpha_j)\right\rangle +\left\langle \sin(\omega t + \alpha_i) \cos(\omega t +
  \alpha_j) \right\rangle =0.
\end{equation}
Furthermore,
\begin{eqnarray}%16
  \left\langle \overrightarrow{D}^2(t) \right\rangle =\sum\limits_{i,j}
  \left\langle \overrightarrow{d}_i(t)
  \cdot \overrightarrow{d}_j(t) \right\rangle = \sum\limits^{N_0}_{i=1}
  \left\langle \overrightarrow{d}_i(t)
  \cdot \overrightarrow{d}_i(t) \right\rangle  + \nonumber \\
 + \sum\limits^{N_0(N_0-1)}_{i\neq j}\left\langle \overrightarrow{d}_i(t)
  \cdot \overrightarrow{d}_j(t) \right\rangle  =e^2 r^2_L \left[
  N_0 +\sum\limits^{N_0(N_0-1)}_{i\neq j}\cos (\alpha_i-\alpha_j)
  \right]
\end{eqnarray}
and
\begin{equation}%17
  \left\langle (\overrightarrow{D}\cdot
  \overrightarrow{n})^2 \right\rangle =e^2r^2_{L}\left[N_0
  +\sum\limits^{N_0(N_0-1)}_{i\neq j}\cos (\alpha_i-\alpha_j)
  \right]  (n^2_x+n^2_y)/2.
\end{equation}
Using the relations
$$
n^2_x + n^2_y + n^2_z=1, \quad
1-\frac{n^2_x+n^2_y}{2}=\frac{1+n^2_z}{2} ,
$$
we get
\begin{equation}%18
  \left\langle dI \right\rangle =\frac{e^2 \omega^2 V^2_{\perp}}{c^3}\left[ N_0
  +\sum\limits^{N_0(N_0-1)}_{i\neq j}\cos (\alpha_i-\alpha_j)
  \right] \frac{1+n^2_z}{2} \frac{d\Omega}{4\pi}.
\end{equation}
The factor $ (1 + n^2_z)/2 $ reflects the anisotropy properties of the dipole
radiation of nonrelativistic electrons. The second term in the square brackets
describes the correlation effects connected with the mutual aligning of
dipoles. If correlations are absent, i.e., if the contributions from all $\cos
(\alpha_i-\alpha_j) $ are mutually compensated and give zero in the sum, then
only the first term in (18) works, which corresponds  to the total radiation of
$N_0 $ independent elementary dipoles. It is easy to describe situations where
the second term in (18) completely compensates the first one so that dipole
radiation does not arise. These are, of course, the situations where the total
dipole moment $ \overrightarrow {D} $ becomes equal to zero. Consider, for
example, the case where $N_0/2 $ dipoles are oriented precisely in one
direction, and the rest $N_0/2 $ dipoles are oriented in the opposite
direction. Then the correlations of pairs in each of these groups are
constructive and give $\cos (\triangle \alpha) =1 $; there are $N_0 (N_0/2-1)/2
$ such pairs in one group and as many in the other, i.e., their common positive
contribution is equal to $N_0 (N_0/2-1) $. Correlations between pairs formed by
dypoles of different groups are destructive and give $\cos  \pi =-1 $ so that
the contribution of such pairs is, obviously, equal to $-2 (N_0/2) ^2 $. Adding
these contributions together with the first term in (18), we obtain: $ N_0 +
N_0 (N_0/2-1)-2 (N_0/2) ^2=0 $, as it should be from the physical viewpoint
because we consider the case with $ \overrightarrow {D} =0 $.

In the case of total correlation, where all $\cos  (\alpha_i-\alpha_j) =1 $,
formula (18) gives
\begin{equation}%19
  \left\langle dI \right\rangle _{corr}=N^2_0 \frac{e^2 \omega^2
  V^2_{\perp}}{c^3} \; \frac{1+n^2_z}{2} \; \frac{d\Omega}{4\pi} \; ,
\end{equation}
i.e., the intensity grows in $N_0 $ times as compared with the radiation of
$N_0 $ uncorrelated dipoles. This is precisely the SR effect [1].

The radiation time of such correlated dipoles radiating coherently will
decrease $N_0 $ times as compared with time (7):
\begin{equation}%20
  \tau_{coh}=\tau / N_0.
\end{equation}

In reality, it is possible to expect only partial positive correlation of
phases, i.e., partial  aligning of all dipoles, for wich the average over
ensemble value of cosine is positive
\begin{equation}%21
\left\langle \cos (\alpha_i-\alpha_j) \right\rangle  \equiv \left\langle \cos
\triangle \alpha \right\rangle   \;  >  0.
\end{equation}
Having replaced all $\cos  (\alpha_i-\alpha_j) $ in (18) by their average
values, we obtain
\begin{equation}%22
\left\langle dI \right\rangle =\frac{e^2 \omega^2
  V^2_{\perp}}{c^3} \; \left[ N_0 + N_0(N_0-1)\left\langle \cos  \triangle
  \alpha \right\rangle   \right] \; \frac{1+n^2_z}{2} \; \frac{d\Omega}{4\pi}.
\end{equation}
In this case, the intensity of coherent radiation is proportional to $N^2_0
\left\langle \cos \triangle\alpha \right\rangle   $.

Now we  proceed  to the consideration of   the mechanism of spontaneous
aligning of the dipoles giving rise to the SR regime.
\bigskip
\begin{center}
\section{Polarization phase transition  in ``coherence domains".}
\end{center}

To solve the problem  of the phase transition, we apply here the Weiss method
of self-consistent mean  field [8] confirmed in the theory of spontaneous
magnetization.  Consider the potential energy of a trial dipole $
\overrightarrow {d}_0 \left( \overrightarrow {r} _0, t \right) $ with the
electric field $ \overrightarrow {E} \left( \overrightarrow {r} _0, t \right) $
induced at a point $ \overrightarrow {r} _0 $ by the rest $ (N_0-1) $ dipoles:
\begin{equation}%23
  U(d_0)=-\overrightarrow{d}_0(\overrightarrow{r}_0,t) \cdot
  \overrightarrow{E}(\overrightarrow{r}_0,t),
\end{equation}
where
\begin{equation}%24
  \overrightarrow{E}(\overrightarrow{r}_0,t)=\sum\limits^{N_0-1}_j
  \frac{3\overrightarrow{n}_j (\overrightarrow{n}_j \cdot
  \overrightarrow{d}_j(t)) -\overrightarrow{d}_j(t)}
  {|\overrightarrow{r}_0-\overrightarrow{r}_j|^3},
\end{equation}
$$
  \overrightarrow{n}_j=(\overrightarrow{r}_0-\overrightarrow{r}_j)
  /|\overrightarrow{r}_0-\overrightarrow{r}_j|.
$$
All dipoles are rotating according to law (4) and radiate, and their electric
field is not static but also is rotating with frequency $ \omega $. Therefore,
the use of expression (24) for the field $ \overrightarrow {E} (t) $ requires
explanation. The point is that conditions (11) that determine the ``coherence
volume" imply that different dipoles of a domain   are in the so-called near
zone $ (R_0 \ll \lambda) $ with respect to each other, where the main term in
the decomposition of the retarded potentials and fields in powers of the small
parameters $ (r _ {L}/R_0) $ and $ (r _ {L} / \lambda) $ turns out to be
precisely expression (24) (in this respect, see, e.g., [9]).

Averaging Eq.~(23) over the rotation period, we  notice that the points $
\overrightarrow {r} _0 $ and $ \overrightarrow {r} _j $ characterize not the
instantaneous positions of the rotating electrons but the positions of the
static centers of rotation and, consequently, do not depend on time. Substiting
(24) into (23) and averaging over the period,  we  obtain
\begin{equation}%25
  \left\langle U(r_0) \right\rangle  =-\frac{d^2_0}{2}\; \sum\limits^{N_0-1}_{j=1}\frac{1-3(n_{jz})^2}
  {|\overrightarrow{r}_0-\overrightarrow{r}_j|^3} \;
  \cos (\alpha_0-\alpha_j),
\end{equation}
where $d_0=er_L$, $n_{jz} = (z_0-z_j) / |\overrightarrow {r}_0 -
\overrightarrow{r}_j|$, and the relation $3[n^2_{jx}+n^2_{jy}]-2=1-3n^2_{jz}$
is used. One can approximate the sum in (25) by  the integral with respect to
the coordinates $r_j $ over the ``coherence volume" $V _ {coh} $ with the
obvious measure $n_e dV_j $ representing the mean number of dipoles in the
volume element $dV_j \equiv (d\overrightarrow {r} _j) $ in the neighborhood of
the point $ \overrightarrow {r} _j $. But, prior to  writing this integral, we
notice that, because the position of our trial dipole $ \overrightarrow {d} _0
(\overrightarrow {r_0}, t) $ can be arbitrary, it is necessary to average (25)
over this parameter, i.e., to introduce the additional integration $
(d\overrightarrow {r_0})/V _ {coh} $. Moreover, in the spirit of the mean-field
method, we replace $\cos (\alpha_0-\alpha_j) $ in (25) by its value $
\left\langle \cos \triangle \alpha \right\rangle  $ averaged over the ensemble.
After all this averaging, we get the expression
\begin{equation}%26
  \left\langle U \right\rangle  =-\frac{d^2_0}{2}\; n_e\left\langle \cos  \triangle \alpha  \right\rangle
   \int (d\overrightarrow{r_0})/V_{coh}\int d\overrightarrow{r}_j
   \frac{1-3(n_{jz})^2}
  {|\overrightarrow{r}_0-\overrightarrow{r}_j|^3}.
\end{equation}
By making the change of variables $ \{
\overrightarrow{r}_0,\overrightarrow{r}_j \} \rightarrow \{
\overrightarrow{r}=\overrightarrow{r}_0-\overrightarrow{r}_j$,
 $\overrightarrow{R}=(\overrightarrow{r}_0+\overrightarrow{r}_j)/2 \}$
and integrating over $d\overrightarrow {R} $, we get
\begin{equation}%27
\left\langle U \right\rangle  =-\frac{d^2_0}{2}\; n_e\left\langle \cos
\triangle \alpha  \right\rangle
   \int\limits_{V_{coh}}\frac{r^2-3z^2}{r^5}(d\overrightarrow{r}).
\end{equation}
Aligning of the dipoles in the same direction is energetically favourable
because of reduction of the negative contribution to potential energy $
\left\langle U \right\rangle $. Therefore,  the energetically preferable
correlations will occur only in that part of the ``coherence volume" in which
the region of integration over the relative coordinates satisfies the condition
\begin{equation}%28
  r^2-3z^2 > 0,
\end{equation}
i.e., in the region similar to a flattened  circular cylinder. We call this
part of the coherence region by ``coherence domain", or by ``domain of
self-polarization". In a similar neighboring domain, the direction of the
average vector of polarization should be close to the opposite one to minimize
the positive energy of the electric field of polarization in the system as a
whole. It is  known that, by similar reasoning, the  macroscopic volumes of
magnetics and ferroelectrics are also divided into domains. Adjacent domains
are separeted by transition regions (``domain walls") within the limits of
which the turning of the polarization vector takes place.

We  return now to the estimation of the integral in (27) with constraint (28).
It is convenient  to calculate it in the cylindrical coordinates $ (\rho, z,
\varphi) $, in which
\begin{equation}%29
  r^2=\rho^2+z^2, \quad r^2-3z^2=\rho^2-2z^2 >0.
\end{equation}
We  consider this integral separately:
\begin{equation}%30
  I(\rho_1,\rho_2)=\int\limits^{2\pi}_0 d\varphi \;
  \int\limits^{\rho_2}_{\rho_1}\rho d \rho \int\limits^{\rho
  /\sqrt{2}}_0 2dz \; \frac{\rho^2-2z^2}{(\rho^2+z^2)^{5/2}}.
\end{equation}
Integration over $z $  in the specified limits gives $2/3\sqrt {3} \rho^2 $
and, as a result,
\begin{equation}%31
I(\rho_1,\rho_2)=\frac{4\pi}{3\sqrt{3}}
\int\limits^{\rho_2}_{\rho_1}\frac{d\rho^2}{\rho^2}=
\frac{4\pi}{3\sqrt{3}}\ln \left( \frac{\rho_2}{\rho_1} \right)^2,
\end{equation}
\begin{equation}%32
\left\langle U \right\rangle  =-\frac{2\pi}{3\sqrt{3}}\ln \left(
\frac{\rho_2}{\rho_1} \right)^2 \cdot d^2_0 n_e \left\langle \cos  \triangle
\alpha \right\rangle  .
\end{equation}

Now we  consider  the question about the minimal and maximal limits $ (\rho_1,
\rho_2) $ of the relative coordinate $ \rho =\sqrt {(x_1-x_2) ^2 + (y_1-y_2)
^2} $ in the plane between two dipoles in the coherence domain. We  remind,
that the characteristic sizes of the initial ``coherence volume" were
determined by conditions (11): $r_L \ll R_0 \ll \lambda $. Inequality (29)
means that, in relative coordinates, ``the coherence domain" has the form of a
flattened circular cylinder with radius $2R_0 $. Hence, the maximal value of $
\rho $ is $ \rho_2=2 R_0 $, and the minimal value of $ \rho_1 $ should be taken
to be about twice the Larmor radius: $ \rho_1 \sim 2r_L $, because, at smaller
distances between dipoles centers, the interaction between a pair of electrons
is not of dipole character any more, and it is impossible  to use the dipole
formulas. It must be noted that, under the conditions considered here, the
radius of the Debay screening is a little bit smaller than $r_L$.  Thus, we can
write $ \ln (\rho_2/\rho_1) \approx \ln (R_0/r_L) $. Because the ratio $R_0/r_L
$ enters under a sign of logarithm, the result is weakly sensitive to the exact
value of this ratio. Thus, taking into account also the condition $R_0 \ll
\lambda $, we can replace $R_0 $ here by the quantity of the order of $
\lambda/10 $. As a result, we can write
\begin{equation}%33
  \ln \left( \frac{\rho_2}{\rho_1} \right)^2\approx
  \ln \left( \frac{\lambda  \omega}{10 V_{\perp}} \right)^2 =
  \ln \left( \frac{2 \pi c}{10 V_{\perp}} \right)^2
 = \ln \left( \frac{4\pi^2 m c^2}{10^2 m V_{\perp}^2}
\right)\approx
   \ln \left( \frac{mc^2}{5E_{\perp} } \right).
\end{equation}
Hence, at this stage, inequalities (11) turn into the condition
\begin{equation}%34
  E_{\perp} \ll mc^2/5 \approx 100 \, \mbox{keV}.
\end{equation}

With the account of (33), expression (32) takes the following form
\begin{equation}%35
\left\langle U \right\rangle  \approx -\frac{2\pi}{3\sqrt{3}}\ln \left(
\frac{mc^2}{5E_{\perp}} \right)
 d^2_0 n_e \left\langle \cos  \triangle \alpha  \right\rangle  .
\end{equation}

To find $ \left\langle  \cos  \triangle \alpha  \right\rangle   $, we turn to
the Weiss method [8]. For this purpose, in the beginning it is necessary to
consider the response of our system of rotating dipoles in the $ (x, y) $ plane
\begin{equation}%36
  \overrightarrow{d_j}=d_0\{ \cos (\omega t + \alpha_j),
  \sin (\omega t + \alpha_j),0\}
\end{equation}
to the external homogeneous electric field
\begin{equation}%37
 \overrightarrow{E_e}=E_e\{ \cos (\omega t + \alpha_0),
  \sin (\omega t + \alpha_0),0\}
\end{equation}
rotating synchronously with dipoles. The potential energy of a
dipole $ \overrightarrow {d_j} $ in this field, averaged over the
period of rotation,  is
\begin{equation}%38
  \overline{U}=-\overline{\overrightarrow{d_j}(t) \cdot
  \overrightarrow{E_e}(t)}=-d_0 E_e \cos (\alpha_j-\alpha_0).
\end{equation}
Thus, one can see that the aligning of dipoles along the field with the
radiation of released energy  is energetically favourable. Thermal fluctuations
suppress this tendency. They occur in rarefied magnetized plasma mainly in the
form of plasma fluctuations and Alfven waves. The distribution over the phase
differences \mbox {$ \triangle \alpha =\alpha_j-\alpha_0 $} is thus given by
the Boltzmann formula
\begin{equation}%39
  \rho(\triangle \alpha)=
  C \; \exp \left(\frac{-U(\triangle \alpha)}{kT} \right)
  = C \; \exp
  \left( \frac{d_0E_e \cos  \triangle \alpha}{kT  } \right)
\end{equation}
with the normalization factor
\begin{equation}%40
  C^{-1}=\int\limits^{2 \pi}_0 d(\triangle \alpha) \;
  \exp \left( \frac{d_0 E_e}{kT} \; \cos \triangle \alpha \right) =
  2 I_0 \left( \frac{d_0 E_e}{kT} \right),
\end{equation}
where $I_0 (x) $ is the modified Bessel function of zero order [10]. The mean
value of $ \left\langle  \cos \triangle \alpha  \right\rangle   $ is determined
by the integral
\begin{eqnarray}
  \left\langle \cos \triangle \alpha \right\rangle  =
  \int\limits^{2 \pi}_0 \cos \triangle \alpha \rho(\triangle \alpha)
  d(\triangle \alpha) &=&
  \frac{d}{dx}\ln\int\limits^{2 \pi}_0 d(\triangle
  \alpha)\exp \{ x(\cos \triangle \alpha) \}  \nonumber \\
  &=& \frac{d}{dx}\ln
  I_0(x)=\frac{I_0'(x)}{I_0(x)}=\frac{I_1(x)}{I_0(x)},
\end{eqnarray}
where
\begin{equation}%42
  x=d_0 E_e/kT.
\end{equation}

So, the external field $E_e $ induces the nonzero correlator (41), i.e., in
other terms, polarizes the system. The measure of polarization is the mean
dipole moment of unit volume
\begin{equation}%43
  P=n_e d_0 \left\langle \cos \triangle \alpha \right\rangle  .
\end{equation}
Polarization generates the additional internal electric field
\begin{equation}%44
  E_p=\nu \cdot P \; ,
\end{equation}
where $ \nu $ is some dimensionless parameter which will be defined below. The
field $E_p $, in turn, strengthens the polarization. This feedback effect will
be taken into account if in (42) we replace $E_e $ by the sum $E_e + E_p=E_e +
\nu P $ in correspondence with the ideology of the self-consistent mean field
developed by Weiss. As a result, we obtain the following nonlinear equation for
the determination of the polarization $P $:
\begin{equation}%45
  P=n_e d_0\left\langle \cos \triangle \alpha \right\rangle  =n_ed_0 F \left( \frac{d_0(E_e+\nu
  P)}{kT}\right),
\end{equation}
where
\begin{equation}%46
  F(x)=I_1(x)/I_0(x).
\end{equation}
By excluding the external field ($E_e\rightarrow 0 $), we obtain ``the
self-consistency equation" for $P$:
\begin{equation}%47
  P=n_ed_0 F \left( \frac{d_0\nu}{kT} P \right).
\end{equation}

We consider now the conditions of  existence  of its nontrivial solutions.
Introducing the variable $z = (P d_0\nu/kT) $, we rewrite equation (47) in the
following form:
\begin{equation}%48
  z=\frac{d_0^2 \: \nu \: n_e}{kT}\; F(z).
\end{equation}
The function $F (z) $ has the asymptotics [10]
\begin{equation}%49
  F(z)=1-\frac{1}{2z}-\frac{1}{8z^2}-\cdots, \quad z \gg 1,
\end{equation}
\begin{equation}%50
  F(z)=\frac{z}{2}\left[ 1-\frac{z}{8}+\cdots \right], \quad z \ll 1.
\end{equation}
From (48) and (49) it is follows that, at large $z $, the solution exists and
corresponds to the polarization of saturation
\begin{equation}%51
  z=\frac{\nu n_e d_0^2}{kT} \gg 1,
\end{equation}
\begin{equation}%52
  P_{max}=n_ed_0=\frac{kT}{\nu d_0}\; z \gg \frac{kT}{\nu d_0}.
\end{equation}
To determine the threshold value of the electron density  above which there
arises a nontrivial solution of equations (47), (48), it is necessary to
consider asymptotic (50). Restricting ourselves to the first term, we obtain
from (48)
\begin{equation}%53
  z=\frac{\nu \: n_e \: d^2_0}{2kT} \; z, \quad z \ll 1.
\end{equation}
It follows from this equation that the critical (threshold) value of the
density $n_e $ is determined by the condition
\begin{equation}%54
  \left( \frac{\nu}{2} \; \frac{n_e d_0^2}{kT} \right)_c =1.
\end{equation}
For $n_e \geq n_{ec}$, nontrivial domain self-polarization appears.

To determine the factor $ \nu $, it is necessary to compare the expression for
potential energy following from (43), (44), and (38),
\begin{equation}%55
\left\langle U \right\rangle  =-\nu n_e d^2_0 \left\langle \cos \triangle
\alpha \right\rangle ,
\end{equation}
with the previously obtained expression (35). Requiring the equality between
them, we obtain
\begin{equation}%56
  \nu = \frac{2\pi}{3\sqrt{3}} \ln \left( \frac{mc^2}{5E_{\perp}} \right),
   \; \;
  \left( E_{\perp} \ll \frac{mc^2}{5} \right).
\end{equation}
As an estimate, for example, taking $E _ {\perp} =1~\mbox {keV} $, we obtain $
\nu \approx 5.56 $.

It is useful to express $d^2_0 $ through the energy $E _ {\perp} $ and magnetic
field $H $:
\begin{equation}%57
  d^2_0=(er_L)^2=\left( \frac{e V_{\perp}}{
  \omega} \right)^2=\left( \frac{mc V_{\perp}}{H} \right)^2=
  \frac{2mc^2 E_{\perp}}{H^2 }.
\end{equation}
As a result, the criterion of the occurrence of domain self-polarization of the
inverted electron system on high Landau levels leading to SR takes the form
\begin{equation}%58
\frac{2\pi}{3\sqrt{3}} \ln \left( \frac{mc^2}{5E_{\perp}} \right)
\frac{mc^2}{H^2}\frac{n_e E_{\perp}}{kT}\geq 1.
\end{equation}
\bigskip

\section{Problem of the power of the decameter radiation of Jupiter.}

As an illustration  of the application of criterion (58), we  consider the
values of parameters in (58) characteristic for the lower magnetosphere of the
active system  Jupiter-Io at the base of the so-called ``Io flux tube", where,
according to the observations, one of the sources of powerful decameter
radiation is located [7]. The characteristic values of the parameters in this
case are: $H\sim 10$~Gs, $T \sim 10^3$~K, and $E _ {\perp} \sim 1$~keV.
Substituting these values into (58), we  find the critical density of inverted
electrons $ (n_e) _c $, above which the criterion (58) will be satisfied: $n_e
\geq (n_e) _c = 2 \cdot 10^3~\mbox{cm} ^ {-3} $. This is quite a modest
requirement to the density of inverted fast electrons at the base of the Io
flux tube which can easily be satisfied, so there are all grounds to believe
that the sporadic super-power decameter radiation of the Jupiter-Io system can
be connected with the generation of SR.

Let us consider this question in more detail. We notice at once that, as in the
SR regime, the energy $E _ {\perp} (t) $ entering (58) quickly decreases, and
the density $n_e $ satisfying criterion (58) at $E _ {\perp} =E _ {\perp} (0) $
can cease to satisfy it with the decrease in $E _ {\perp} (t) $. Therefore, for
an effective and sufficiently long operating SR regime, one obviously requires
a large enough excess of $n_e $ over $n _ {ec} \sim 2 \cdot 10^3 $~cm$ ^ {-3}
$. Thus, we conclude that, at the transition to SR regime, electrons at the
base of the Io flux tube near Jupiter's magnetic pole  must have density $n_e
\geq 10^4 $~cm$ ^ {-3} $.

According to the basic assumption [6,7] now accepted, the streams of energetic
electrons appear near  Jupiter magnetic poles  due to their acceleration up to
several keV in the Io ionosphere  and further movement to the Jupiter along the
lines of the magnetic dipole fields. Acceleration of electrons is accounted by
the induction of electromotive force $ \sim 400 $~kV in the Io body and
ionosphere due to the Io motion through the Jupiter's magnetic field. The lines
of the magnetic field of the Io flux tube converge to Jupiter's magnetic poles.
So, the initial area of acceleration of electrons in the Io ionosphere $S_0
\sim 10^{16}$--$10^{17}$~cm$^2 $ will be ``projected" by the magnetic lines to
an area smaller approximately by two orders of magnitude, $S_1 \sim 10 ^ {-2}
\cdot S_0 \sim 10 ^ {14}$--$10^{15}$~cm $ ^ 2 $ near Jupiter's poles.
Therefore, the supercritical density of inverted fast electrons $n_e \geq
10^4$~cm$ ^ {-3} $, necessary for the occurrence of SR, will be provided by the
initial density $n _0 \geq 10^2$~cm$ ^ {-3} $ near Io. It is quite an
acceptable value in view of the known data about Io and its ionosphere [11].

Electrons moving along the Io flux tube initially have density below the
critical value and, therefore, radiate rather weakly. But, near the Jupiter
magnetic poles, the density of electrons  reachs the supercritical value, and
they pass to the SR regime. Thus, in this region, the intensity of collective
cyclotron radiation grows approximately by ten orders of magnitude ($N_0=n_e
\cdot V _ {coh} \sim 10 ^ {10} $), what accounts for the observable huge power
of decametric radiation from a rather small volume of the circumpolar area
occupied by the source [7]. A separate work will be devoted to more detailed
calculation of this power and also to the interpretation of the rather unusual
dynamical spectra of the decameter radiation of the Jupiter-Io system within
the framework of the present model.

\section*{Acknowledgments}

This work  is supported in part by the National scientific fund of Switzerland
in the frameworks of SCOPES.

\end{document}